\documentclass[journal]{IEEEtran}

\usepackage{url} 
\usepackage{graphics} 
\usepackage{amsmath,amsfonts,amssymb,epsfig} 
\usepackage{lscape} 
\usepackage{latexsym}

\newtheorem{theorem}{\bf Theorem}[section] 
\newtheorem{lemma}[theorem]{\bf Lemma}

\newtheorem{definition}[theorem]{\sc Definition} 
 
\newenvironment{remark}{{\noindent \bf Remark.}}{} 

\begin{document}

\title{Bounds on Key Appearance Equivocation for Substitution Ciphers}

\author{Yuri Borissov and~Moon Ho Lee,~\IEEEmembership{Senior Member,~IEEE}
\thanks{Yuri Borissov is with the Institute of Mathematics and Informatics, 
Bulgarian Academy of Sciences, Sofia 1113, Bulgaria. Moon Ho Lee is with the Institute of Information and Communication,
Chonbuk National University, Jeonju 561-756, R. Korea.
}} 

\maketitle

\begin{abstract}\noindent
The average conditional entropy of the key given
the message and its corresponding cryptogram, $H({\bf K} \vert {\bf M},{\bf C})$,
which is reffer as a key appearance equivocation, was proposed as a theoretical 
measure of the strength of the cipher system under a known-plaintext attack by Dunham in 1980.
In the same work (among other things), lower and upper bounds for $H({\cal S}_{\cal M} \vert {\bf M}^L {\bf C}^L)$
are found and its asymptotic behaviour as a function of cryptogram length $L$ is described for simple substitution ciphers
i.e. when the key space ${\cal S}_{\cal M}$ is the symmetric group acting on a 
discrete alphabet ${\cal M}$. In the present paper we consider the same problem 
when the key space is an arbitrary subgroup ${\cal K} \triangleleft {\cal S}_{\cal M}$ and generalize
Dunham's result.
\end{abstract}

\begin{keywords} 
key appearance equivocation, substituion ciphers. 
\end{keywords} 

\section{Introduction}
Shannon in his seminal paper \cite{Sha} showed that the conditional entropies of the key
and message given the cryptogram can be used as a theoretical measure of strength of the 
cipher system when assuming unlimited cryptanalytic computational capabilities. These conditional
entropies are called the key and message equivocation, respectively.

In general it is diffucult to calculate these equivocations explicitly. For that Shannon
established in \cite{Sha} a general lower bound and introduced a random cipher model which
would approximate the behaviour of complex practical ciphers. Afterward, Hellman \cite{Hel} 
reviewed and extended Shannon's information-theoretic approach and showed that
random cipher model is conservative in that a randomly chosen cipher is essentially the worst possible.
Later on Blom \cite{Bl} obtained exponentially tight bounds on the key equivocation for simple
substitution ciphers. In \cite{JGD} to derive  bounds for simple substitution ciphers on the
message equivocation in terms of the key equivocation, Dunham derived such bounds for so-called key
appearance equivocation. This author pointed out also, that it can be considered as a theoretical measure of
the strength of the cipher system under known-plaintext attack. Another contribution of this subject is the Sgarro's work \cite{Sgarro}.

In Section II we give the necessary background and state a theorem which gives the bounds on
the key appearance equivocation for substitution ciphers when the key space is confined to a subgroup ${\cal K}$
of the group ${\cal S}_{\cal M}$ of all substitutions of a discrete alphabet ${\cal M}$.
In Section III we discuss four applications of the stated theorem in some particular cases.
Finally, we conclude in Section IV.
 
\section{Lower and Upper Bounds for the Key Appearance Equivocation}
For basic definitions and notions we reffer to \cite{Sha},\cite{JGD} and \cite{Ga}. 
Let a memoryless message source with a discrete finite alphabet ${\cal M} = \{1,2,\ldots,N\}$ be given. 
The probability of a symbol $n$ is denoted by $P_{\cal M}(n)$. The cryptogram alphabet ${\cal C}$ is taken to be 
the same as ${\cal M}$, and the key space is ${\cal K} \triangleleft {\cal S}_{\cal M}$ -- an arbitrary subgroup of the 
the symmetric group acting on ${\cal M}$. For every $\pi \in {\cal K}$ the cryptographic transformation 
${\rm T}_{\pi} : {\cal M}^L \rightarrow  {\cal M}^L$ is determined in the following way: 
If ${\bf  m}^L = m_1 m_2 \ldots m_{L}$ is a message of length $L$, then the cryptogram is 
${\bf c}^L = {\rm T}_{\pi}({\bf m}^L) \stackrel{\rm def}{=} \pi(m_1) \pi(m_2) \ldots \pi(m_{L})$.
We assume also that the key and message sources are independent, and the keys are equiprobable, i.e. $P_{\cal K}(\pi)= 1/|\cal K|$.

We make use of the following lemma:

\begin{lemma}\label{Stabilizer}
Let $G$ be a group of substitutions of the finite set $X$. If the set $G(i,j) = \{\pi \in G/\pi(i)=j\}$, where $i$ and $j$ are some 
fixed elements of $X$, is nonempty, then it is a left coset by the stabilizer $St(i)\stackrel{\rm def}{=}\{\tau \in G/ \tau(i) = i\}$.
\end{lemma}
\begin{proof}
Obviously, if $\pi(i) \in G(i,j)$ then for any $\alpha \in St(i)$ we have $\pi \circ \alpha(i) = \pi(i) = j$. Conversely, if $\pi(i) = j$ and $\tau(i) = j$ then
${\pi}^{-1} \circ \tau(i) =  {\pi}^{-1}(j)= i$ hence ${\pi}^{-1} \circ \tau \in St(i)$.
\end{proof}

In order to state the main theorem we need the following definitions:

\begin{definition}
The set $F(\pi) \stackrel{\rm def}{=} \{j/\pi (j) = j \}$ is called a fixed set of $\pi \in \cal {K}$.
\end{definition}

Let us denote by $\cal {K}^{*}$ the set of all substitutions in $\cal {K}$ excluding the identity.
\begin{definition}
The key $\pi \in \cal {K}^{*}$ is called maximal when its fixed set $F(\pi)$ is maximal in sense of inclusion among the sets $F(\tau)$, $\tau \in \cal {K}^{*}$.
\end{definition}
We will denote by ${\cal K_{\rm max}}$ the set of all maximal keys
and for any $\pi \in {\cal K_{\rm max}}$ by $P_{\pi}$ the sum of probablities $\sum_{j \in F(\pi)} P_{\cal M}(j)$.

For completeness of exposition we recall the defintion of key appearance equivocation:
\begin{definition}\label{kae}
\begin{eqnarray*}
H({\bf K} \vert {\bf M}^L {\bf C}^L) \stackrel{\rm def}{=}\! \sum_{{\bf  m}^L \in {\bf M}^L}\!\sum_{{\bf  c}^L \in {\bf C}^L}H({\bf K} \vert {\bf m}^L {\bf c}^L)P_{{\cal M}^L {\cal C}^L}({\bf m}^L{\bf c}^L)
\end{eqnarray*}
and
\begin{eqnarray*}
H({\bf K} \vert {\bf m}^L {\bf c}^L) \stackrel{\rm def}{=} \sum_{k: {\rm T_{k}}({\bf m}^L) = {\bf c}^L} P_{\bf K}(k)log(1/P_{\bf K}(k))
\end{eqnarray*}

\end{definition}

The following theorem is a generalization of the result obtained in \cite{JGD} on the behaviour of key appearance equivocation for simple substitution ciphers 
as a function of cryptogram length.

\begin{theorem}\label{Th1}
Under the above impossed assumptions, let ${\cal K_{\rm max}}$ is nonempty and $R = \max \{P_{\tau}/ \tau \in {\cal K_{\rm max}} \}$. Then  the 
following inequalities hold:
\begin{eqnarray*}
\log(2)R^L \leq H({\cal K} \vert {\bf M}^L {\bf C}^L) \leq \log(\vert{\cal K}\vert)  \vert {\cal K_{\rm max}} \vert  R^L
\end{eqnarray*}
\begin{eqnarray*}
\end{eqnarray*}
\end{theorem}
\begin{remark}
The logarithms are taken for an arbitrary fixed base depending on the unit of entropy measurement. 
\end{remark}

\begin{proof}
Starting from definition of conditional entropy, using the fact that the keys are equiprobable and applying Lemma\ref{Stabilizer}
we consecutively get:
\begin{eqnarray*}
\sum_{{\bf  m}^L \in {\bf M}^L}\!\sum_{{\bf  c}^L \in {\bf C}^L}H({\bf K} \vert {\bf m}^L {\bf c}^L)P_{{\cal M}^L {\cal C}^L}({\bf m}^L{\bf c}^L) =
\end{eqnarray*}
\begin{eqnarray*}
\sum_{{\bf  m}^L \in {\bf M}^L}\!\sum_{{\bf  c}^L \in {\bf C}^L} log(|St({\bf m}^L)|)P_{{\cal M}^L {\cal C}^L}({\bf m}^L{\bf c}^L) = 
\end{eqnarray*}
\begin{eqnarray*}
\sum_{{\bf  m}^L \in {\bf M}^L}log(|St({\bf m}^L)|)\sum_{{\bf  c}^L \in {\bf C}^L}P_{{\cal M}^L {\cal C}^L}({\bf m}^L{\bf c}^L) = 
\end{eqnarray*}
\begin{eqnarray*}
\sum_{{\bf  m}^L \in {\bf M}^L}log(|St({\bf m}^L)|)P_{{\cal M}^L}({\bf m}^L),
\end{eqnarray*}
where $St({\bf m}^L)=\{{\rm T}_{\pi}/{\rm T}_{\pi}({\bf m}^L)= {\bf m}^L, {\pi} \in {\cal K}\}$ is the stabilizer of message ${\bf m}^L$.

Clearly, if $St({\bf m}^L) \not = \{e\}$, where $e$ is identity, we have: $2 \leq |St({\bf m}^L)| \leq |{\cal K}|$. 
Thus the following inequalities hold:
\begin{eqnarray*}
log(2)\sum_{{\bf m}^L:St({\bf m}^L) \not = \{e\}}P_{{\cal M}^L}({\bf m}^L) \leq H({\cal K} \vert {\bf M}^L {\bf C}^L) \leq 
\end{eqnarray*}
\begin{eqnarray*}
log(|{\cal K}|)\sum_{{\bf m}^L:St({\bf m}^L) \not = \{e\}}P_{{\cal M}^L}({\bf m}^L)\;\;\;\;\;\;\;(1)
\end{eqnarray*}

The fact that the message source is memoryless implies for any $\Omega \subset {\cal M}$ and $(m_1m_2 \ldots m_{L}) = {\bf m}^L \in \Omega^{L}$
\begin{eqnarray*}
\sum_{{\bf m}^L}P_{{\cal M}^L}({\bf m}^L) = \sum_{{\bf m}^L}\prod_{l=1}^{L}P_{{\cal M}}(m_{l})
\end{eqnarray*}
\begin{eqnarray*}
 = (\sum_{m \in \Omega}P_{{\cal M}}(m))^{L}
\end{eqnarray*}
Let $R=P_{\pi}$. Since ${\rm T}_{\pi} \in St({\bf m}^L)$ for any ${\bf m}^L \in [F(\pi)]^{L}$ then $[F(\pi)]^{L} \subset \{{\bf m}^L \in {\cal M}^L/St({\bf m}^L) \not = \{e\}\}$.
Therefore the following inequality holds:
\begin{eqnarray*}
R^{L} = (\sum_{m \in F(\pi)}P_{{\cal M}}(m))^{L} = \sum_{{\bf m}^L \in [F(\pi)]^{L}}P_{{\cal M}^L}({\bf m}^L) \leq
\end{eqnarray*}
\begin{eqnarray*}
\sum_{{\bf m}^L:St({\bf m}^L) \not = \{e\}}P_{{\cal M}^L}({\bf m}^L)\;\;\;\;\;\;\;(2)
\end{eqnarray*}

On the other hand, if for some ${\bf m}^L,St({\bf m}^L) \not = \{e\}$ holds, then there exists a maximal key $\tau$ such that ${\bf m}^L \in [F(\tau)]^{L}$.
Therefore we have:
\begin{eqnarray*}
\sum_{{\bf m}^L:St({\bf m}^L) \not = \{e\}}P_{{\cal M}^L}({\bf m}^L) \leq 
\end{eqnarray*}
\begin{eqnarray*}
\sum_{\tau \in {\cal K_{\rm max}}}\sum_{{\bf m}^L \in [F(\tau)]^{L}}P_{{\cal M}^L}({\bf m}^L) = 
\end{eqnarray*}
\begin{eqnarray*}
\sum_{\tau \in {\cal K_{\rm max}}}(P_{\tau})^L \leq \vert{\cal K_{\rm max}}\vert  R^{L}\;\;\;\;\;\;\;(3)
\end{eqnarray*}

From $(2)$ and $(3)$ substituting in $(1)$, we finally obtain:
\begin{eqnarray*}
\log(2)R^L \leq H({\cal K} \vert {\bf M}^L {\bf C}^L) \leq \log(\vert{\cal K}\vert)  \vert{\cal K_{\rm max}}\vert  R^L
\end{eqnarray*}
which is the desired result.
\end{proof}

Note that Theorem\ref{Th1} shows the asymptotic tight exponential behaviour of $H({\cal K} \vert {\bf M}^L {\bf C}^L)$ with exponent base $R$ equal to   
the maximum among sums of symbol probabilities of the fixed sets of maximal keys.

\section{Applications}
We shall consider four applications of Theorem\ref{Th1}. For the first two applications we assume without loss of generality that $P_{\cal M}(1) \geq P_{\cal M}(2) \geq \ldots \geq P_{\cal M}(N)$.

1. Let ${\cal K} = {\cal S}_{\cal M}$ -- the case of simple substitution cipher. Clearly, maximal keys are the transpositions.
Therefore, 
$R_1 = \sum_{j=1}^{N-2}P_{\cal M}(j) = 1 - P_{\cal M}(N) - P_{\cal M}(N-1)$, $|{\cal K}| = N!$ and $\vert{\cal K_{\rm max}}\vert = {N\choose 2}$. This result is obtained in \cite{JGD}.

2. let ${\cal K} = {\cal A}_{\cal M}$, where ${\cal A}_{\cal M}$ is the alternating group acting on ${\cal M}$. It can be easily seen that maximal keys are
the substitutions which can be represented as a superposition of cycle of length $3$ and disjoint to this cycle identity substitution. 
Clearly, these substitutions belong to ${\cal A}^{*}_{\cal M}$. Proceeding as in the previous case we get 
$R_2 = \sum_{j=1}^{N-3}P_{\cal M}(j) = 1 - P_{\cal M}(N) - P_{\cal M}(N-1) - P_{\cal M}(N-2)$, $|{\cal K}| = N!/2$ and $\vert{\cal K_{\rm max}}\vert = {N\choose 3}$. 

3. Let $d$ be a positive integer. We will consider messages of length $L = k  d, k \geq 1$. Since the message source is memoryless it is memoryless also over 
the cartesian product ${\cal M}^d 
$ considered as an alphabet. 

Let $\pi \in {\cal S}_{\Delta}$, where $\Delta = \{1,2, \ldots, d\}$. Define a mapping $\rm T_{\pi}: {\cal M}^d \rightarrow {\cal M}^d$ as 
$\rm T_{\pi}(m_1m_2 \ldots m_d) \stackrel{\rm def}{=} m_{\pi(1)}m_{\pi(2)} \ldots m_{\pi(d)}$. Since $\pi$ is a substitution, it follows that $\rm T_{\pi}$ is a substitution 
of ${\cal M}^d$. The set $\{\rm T_{\pi}/ \pi \in {\cal S}_{\Delta}\}$ with superposition operation is a group isomorphic to ${\cal S}_\Delta$ and it is a subgroup of ${\cal S}_{{\cal M}^d}$.

Furthermore it is well known that any $\pi \in {\cal S}_{\Delta}^{*}$ can be represented as a superposition of disjoint cycles in a unique way to the order of multipliers.
A partition of $\Delta$ corresponds to this representation and it is not dificult to see that the fixed set $F(\rm T_{\pi})$ consists of exactly those ${\bf m}^d \in {\cal M}^d$ whose
letters in numbered places belonging to the same subset of the partition of $\Delta$, coincide. Therefore, if we take $\rho \in {\cal S}_{\Delta}^{*}$ different from $\pi$  such that the partition
of $\Delta$ detrmined by $\rho$ is "more detailed", then the inclusion $F(\rm T_{\pi}) \subset F(\rm T_{\rho})$ holds. The latter shows that those $\rm T_{\pi}$ are maximal for which $\pi$ is represented as
a superposition of one cycle of length $2$ and disjoint to this cycle identity substitution, i.e. $\pi$ is a transposition. 

Taking into account the above considerations it can be easily computed the rate $R_3 = \sum_{j=1}^{N}{P^{2}_{{\cal M}}(j)}$, the order of subgroup $|{\cal K}| = d!$ and the number of the maximal keys 
$\vert{\cal K_{\rm max}}\vert = {d\choose 2}$ for this case. Finally, we note that inequalities of Theorem\ref{Th1} now become: 
\begin{eqnarray*}
\log(2)R_{3}^k \leq H({\cal K} \vert {\bf M}^{kd} {\bf C}^{kd}) \leq \log(d!){d\choose 2}R_{3}^k
\end{eqnarray*}

4. Let now, the alphabet ${\cal M}$ be a finite field with $|{\cal M}|= N$, where $N$ is a power of prime number.  Let ${\cal K}$ be the group of affine transformations
\begin{eqnarray*}  
g: y = ax+b;\;\;\;\;\;\;\; a,b \in {\cal M}, a \not = 0
\end{eqnarray*}
Obviously, each affine transformation $y = ax+b, a \not = 1$ possesses just one fixed point $x_{f} = b/(1-a)$ and when
$b$ runs through ${\cal M}$ the same does $x_{f}$. Moreover translations $y = x+b, b \not = 0$ do not possess any fixed points. Thus, we have 
$R_4 = \max \{P_{\cal M}(n)/ n \in {\cal M}\}$, $|{\cal K}| = N(N-1)$ and $\vert{\cal K_{\rm max}}\vert = N(N-2)$.

\section{Conclusions}
Despite that during the past three decades mainly computational aspects of cryptology have been developped, there is still place for information-theoretic investigations.
An example in this direction is the theorem from the present paper which justifies mathematically the intuitive understanding that the recovery
of the key in known-plaintext attack on substitution ciphers is more difficult when this key possesses many fixed points.

\section{Acknowledgment}
This research was supported in part by Ministry of Information and Communication (MIC) Korea under the IT Foreign Specialist Inviting Program (ITSIP), ITSOC,  
International Cooperative Research by Ministry of Science and Technology, KOTEF, and 2nd stage Brain Korea 21.

\end{document}